\documentclass[pra,twocolumn,superscriptaddress,showpacs]{revtex4-1}

\usepackage{graphics}           
\usepackage{bm}                 
\usepackage{amsmath}            
\usepackage{amssymb}
\usepackage{verbatim}           
\usepackage{amsthm}             
\theoremstyle{plain}            

\def\bra#1{{\langle#1|}}
\def\ket#1{{|#1\rangle}}

\def\tr{{\rm Tr}}

\begin{document}

\title{Discrete spacetime, quantum walks and relativistic wave equations}

\author{Leonard \surname{Mlodinow}}\email{lmlodinow@gmail.com}
\author{Todd A. \surname{Brun}}\email{tbrun@usc.edu}
\affiliation{Center for Quantum Information Science and Technology, University of Southern California, Los Angeles, California}

\date{\today}

\begin{abstract}
It has been observed that quantum walks on regular lattices can give rise to wave equations for relativistic particles in the continuum limit. In this paper we define the 3D walk as a product of three coined one-dimensional walks. The factor corresponding to each one-dimensional walk involves two projection operators that act on an internal coin space; each projector is associated with either the ``forward'' or ``backward'' direction in that physical dimension. We show that the simple requirement that there is no preferred axis or direction along an axis---that is, that the walk be symmetric under parity transformations and rotations that swap the axes of the cubic lattice---leads to the requirement that the continuum limit of the walk is fully Lorentz invariant. We show further that, in the case of a massive particle, this simple symmetry requirement necessitates that inclusion of antimatter---the use of a four-dimensional internal space---and that the ``coin flip'' operation is generated by the parity transformation on the internal coin space, while the differences of the projection operators associated to each dimension must all anticommute. Finally, we discuss the leading correction to the continuum limit, and the possibility of distinguishing through experiment between the discrete random walk and the continuum-based Dirac equation as a description of fermion dynamics. 

\end{abstract}

\pacs{}

\maketitle

\section{Introduction}

Quantum walks \cite{AharonovY93,Ambainis01,AharonovD01,Kempe03} provide a discrete model of particle dynamics. In such a walk, a particle may be located at any of the vertices of a graph or lattice, and its state evolves in a sequence of discrete time steps. In each step, the particle may move to one of the neighboring vertices (connected by an edge). That time evolution is given by a unitary transformation. The particle may also have an internal space, or ``coin'' space, that participates in the unitary evolution. In some cases (such as the quantum walk on the line), this internal space is necessary for nontrivial evolution. 

Quantum walks on general graphs have played a significant role in the development of new quantum algorithms \cite{Childs02,Shenvi03,Ambainis03,Ambainis07,Farhi08,Ambainis10,Reichardt12}. In addition they have been studied on their own account, as relatively simple quantum systems exhibiting a wide range of interesting quantum phenomena: wave propagation, localization due to symmetry breaking, and transitions between quantum and classical behavior, to give just a few examples \cite{Moore02,Yin08,Brun03a,Brun03b,Kendon03,Brun03c,Kendon06}. 

Quantum walks on lattices have also served as models for particles propagating in space, thus providing a link between information theory and relativistic equations for fermion dynamics. For example, it has been shown a number of times, from several points of view, that a suitable quantum walk can approach, in the continuum limit, a relativistic wave equation such as the Weyl or Dirac equations \cite{Bialynicki94,Meyer96,Strauch06,Bracken07,Chandrashekar10,DAriano14,Arrighi14,Chandrashekar13,DAriano15,Succi15}. From this observation, a fascinating literature has arisen that examines how the properties of a quantum walk translate, in that limit, to the properties of fermions \cite{Farrelly14,Bisio15,Arrighi15,Bisio16,Arrighi16,DAriano17,DAriano17b}, or vice versa \cite{Arrighi14,Succi15}.  In particular, it has been shown that in the continuum limit the ratio of the lattice spacing to the time step corresponds to the speed of light; the coin space becomes a two-dimensional internal spin space; and the coin flip operator is the origin of the fermionic mass term. In what follows we will show, in addition, that the requirement that the dynamics be symmetric under any swapping or inversion of the axes leads to existence of antimatter and a continuum limit that is fully Lorentz invariant, and that the generator of the coin flip operator corresponds to the parity operator. 

In many prior derivations, once the form of the quantum walk evolution has been defined, the imposition of unitarity leads to a set of difficult constraint equations. In this paper we use a different construction, taking a form of quantum walk on a lattice that is a product of one-dimensional unitary walks, each involving two projection operators, associated to the ``forward'' and ``backward'' directions corresponding to that dimension. This form, which was studied in \cite{Chandrashekar13,Chandrashekar11}, guarantees that the walk is unitary and has nontrivial dynamics in two or three spatial dimensions. 

Derivations of the Dirac equation in three dimensions have generally started with a two-dimensional internal coin space, which leads to the Weyl equation in the continuum limit. The four-dimensional internal space of the Dirac equation can then be obtained by coupling or ``pasting together'' two copies of the walk. In this paper, we instead invoke discrete symmetries, which require that the internal coin space be (at least) four-dimensional, and leads naturally to the definition of the Dirac gamma matrices.

Symmetry under the swapping or inversion of axes in this walk amounts to two discrete symmetries: invariance under parity transformations, and a ``noncorrelation'' condition (which can be seen as a discrete rotational invariance). We show that, together, these two conditions imply that the differences of the projection operators in each dimension must anticommute. These anticommuting operators are what play the role of the Dirac equation gamma matrices in the continuum limit. This is why the continuum limit is symmetric under rotations and boosts (even though the underlying quantum walk is not), with the dynamics governed by the Dirac equation with nonzero mass and a four-dimensional internal space. 

The rotational symmetry in the continuum (long wavelength) limit means that, in that limit, the underlying lattice is masked from our perception. Because our experimental abilities are limited, if the lattice and time spacing are small enough, the discrete walk and continuous Dirac equation are indistinguishable. Experiment thus cannot rule out the possibility that space-time is discrete---it can only place upper limits on its granularity. On the other hand, because the quantum walk on the lattice does have preferred directions---the lattice axes---these could manifest as small corrections to the Dirac equation at high energies. We derive these corrections, and estimate their magnitude. In principle, a highly sensitive Michelson-Morley type of interferometric experiment could detect them, though the precision required for a lattice spacing at the Planck scale is beyond what can currently be detected.

The structure of the paper is as follows. In Sec. II we briefly review the mathematics of coined quantum walks. In Sec. III we introduce our approach in the context of a single space dimension, and introduce the internal space projection operators that are associated with the forward and backward walk directions. In Sec. IV we generalize the walk to three space dimensions. In Sec. V we show that the parity operator and the differences of the projection operators associated with each space dimension must all anticommute. Then we introduce an {\it equal-norm condition} as a noncorrelation condition on the 3D walk, and take the continuum limit. In Sec. VI we calculate the leading correction to the continuum limit and discuss the possibility of distinguishing through experiment whether a discrete random walk or the continuum Dirac equation is the correct description of fermion dynamics. 

\section{Review of Quantum Walks}

We now briefly review the most common formulation of discrete-time quantum walks.  (Since this paper only concerns discrete-time walks, we will henceforward refer to them simply as quantum walks or walks.)

The state space of a quantum walk comprises two parts:  a position space, and an internal space, often referred to as the ``coin'' space.  The allowed positions are identified with the vertices of a graph $G$; the edges of the graph connect neighboring vertices.  Two vertices are neighbors if it is possible for the particle (or ``walker'') to move from one to the other in one timestep.  In the current work, we will consider only undirected graphs; so if it is possible to move from vertex $v_1$ to vertex $v_2$, it is also possible to move the other way.

The evolution of a quantum walk is given by a time evolution unitary, which generally takes the following form:
\begin{equation}
\ket{\psi_{t+1}} = U\ket{\psi_t} = \left( \sum_j S_j \otimes P_j \right) \left( I\otimes C \right)\ket{\psi_t} ,
\label{eq:QWalkEvolution}
\end{equation}
where the $\{S_j\}$ are {\it shift operators} that move the particle from its current position to its neighbor in the direction $j$; the $\{P_j\}$ are orthogonal projectors on the internal space; and $C$ is a unitary that acts on the internal space, often called the ``coin flip'' unitary.  The idea is by analogy to a series of coin flips.  The projectors $\{P_j\}$ correspond to different faces of the coin, which indicate which direction to move; the unitary $C$ scrambles the faces, so that one does not constantly move in the same direction.  But in the unitary case, unlike classical random walks, the evolution is always invertible: interference effects play a profound role in the evolution.

In the course of this work, we will generalize this basic quantum walk in various ways, but the state space and evolution unitary will essentially retain this structure.

One of the simplest examples is the quantum walk on the line.  The position space has a basis $\{\ket{x}\}$ corresponding to the integers $x\in\mathbb{Z}$, and the coin space is two-dimensional, with basis vectors labeled $\{ \ket{R},\ket{L} \}$.  The evolution operator becomes
\begin{equation}
U = \left( S \otimes \ket{R}\bra{R} + S^\dagger \otimes \ket{L}\bra{L} \right) \left( I\otimes C \right) ,
\label{eq:OneDWalk}
\end{equation}
where the shift operator is $S\ket{x} = \ket{x+1}$ and $C$ is a $2\times2$ unitary matrix.  Various coin-flip matrices have been studied for this walk; the canonical example is the Hadamard coin flip,
\begin{equation}
C = \frac{1}{\sqrt2} \left(\begin{array}{cc} 1 & 1 \\ 1 & -1 \end{array} \right) .
\end{equation}
In this paper, we will generally consider families of coin-flip unitaries, parametrized as rotations in Hilbert space:
\begin{equation}
C(\theta) = e^{-i \theta Q } ,
\end{equation}
where $Q$ is a Hermitian operator that acts on the internal space.

Henceforth, we will suppress the tensor product symbol $\otimes$ for compactness unless it is absolutely necessary to be clear.  One can think of this by replacing operators that act only on one of the component of the Hilbert space with an operator that acts as the identity on the other component; e.g., $S \rightarrow S\otimes I$, $C(\theta) \rightarrow I\otimes C(\theta)$, and so forth.

\section{The Dirac Equation in 1D}

\subsection{The Continuum Limit of a Quantum Walk}

We will illustrate some aspects of our method employing the simple case of the 1D quantum walk whose evolution operator is given in Eq.~(\ref{eq:OneDWalk}). What would constitute a well-defined continuum limit? In this limit, the discrete structure of both time and space must be well-approximated by continuous functions. Let us introduce a distance scale $\Delta x$ between positions on the line, and a time scale $\Delta t$ for a single step of the quantum walk. We assume that the amplitudes, $\psi(j\Delta x)$, of the wavefunction $\ket{\psi}= \sum_{j=-\infty}^\infty \psi(j\Delta x) \ket{j\Delta x}$ varies slowly as a function of the position $x=j\Delta x$, and that the time evolution of the state in a single step of the walk is also small. Then we can describe the actions of the shift operator and the time evolution operator by
\[
S\ket{x} = \ket{x+\Delta x} , \ \ \ U\ket{\psi(t)} = \ket{\psi(t+\Delta t)} ,
\]
where the allowed positions and times are integer multiples of $\Delta x$ and $\Delta t$, respectively.  We can now define {\it difference operators}:
\begin{equation}
\partial_x = \frac{S - S^\dagger}{2\Delta x} , \ \ \ 
\partial_t = \frac{U - I}{\Delta t} .
\end{equation}
In the continuum limit, these difference operators will become derivatives.

We now rewrite the quantum walk evolution Eq.~(\ref{eq:QWalkEvolution}) for the 1D case in terms of these difference operators:
\begin{eqnarray}
\ket{\psi(t+\Delta t)} &=& (\Delta t \partial_t + I)\ket{\psi(t)} \nonumber\\
&=& \left( (2\Delta x \partial_x + S^\dagger)P_R
  -  (2\Delta x \partial_x - S)P_L \right)  \nonumber\\
&& \times e^{-i\theta Q} \ket{\psi(t)} .
\label{eq:Cont1D-1}
\end{eqnarray}
In the above equation, $P_R = \ket{R}\bra{R}$ and $P_L = \ket{L}\bra{L}$, and $P_R+P_L = I$.  For a two-dimensional internal space, we can without loss of generality take the Hermitian operator $Q$ to have eigenvalues $\pm 1$, so $Q^2 = I$; we can make this assumption because the trace of $Q$ has no effect except to add a global phase, and the operator norm of $Q$ can be normalized to 1 by absorbing a constant into the parameter $\theta$.  (Later we will argue that this is a natural choice even with a higher-dimensional internal space.)

Now let us rearrange Eq.~(\ref{eq:Cont1D-1}) as follows:
\begin{equation}
\Delta t \partial_t \ket{\psi} = \left( 2\Delta x(\Delta P) \partial_x + (SP_L + S^\dagger P_R) \right) e^{-i\theta Q} \ket{\psi} - \ket\psi ,
\label{eq:Cont1D-2}
\end{equation}
where we've suppressed the time argument $t$ of the wavefunction, and defined a new operator $\Delta P = P_R - P_L$.  This operator also has eigenvalues $\pm 1$, such that $(\Delta P)^2 = I$.

For the continuum limit, we need to satisfy a number of conditions.  First, the parameter $\theta$ must be small, $|\theta| \ll 1$.  Otherwise the wavefunction will necessarily change appreciably in a single timestep.  This means that we can expand
\[
e^{-i\theta Q} \approx 1 - i\theta Q .
\]
Next, the wavefunction $\ket{\psi}$ must change slowly with $x$, so that one can legitimately expand around $x$ in powers of $\Delta x$, i.e., $(S + S^\dagger) \ket{\psi} = 2\ket{\psi} +  O(\Delta x^2) $.  This means that
\[
\left(S P_L + S^\dagger P_R  - I\right) \ket\psi = - \Delta x(\Delta P)\partial_x\ket\psi + O(\Delta x^2) .
\]

Employing this, and keeping terms in Eq.~(\ref{eq:Cont1D-2}) only up to first order in $\Delta t$, $\Delta x$, and $\theta$, and dividing through by $\Delta t$, we get
\begin{eqnarray}
\partial_t \ket{\psi} &=& \left( \frac{\Delta x}{\Delta t} \Delta P \partial_x - i\frac{\theta}{\Delta t} Q \right) \ket{\psi} \nonumber\\
&\equiv& \left( c \Delta P \partial_x - imc^2 Q \right) \ket{\psi} ,
\label{eq:Cont1D-3}
\end{eqnarray}
where we have introduced the parameters $c \equiv \Delta x/\Delta t$ and $mc^2 \equiv \theta/\Delta t$.  Henceforth we will assume a choice of units such that $c = 1$.

\subsection{Operators on the Internal Space}

For a quantum walk with a well-defined continuum limit, we have introduced the two operators $\Delta P = P_R - P_L$ and $Q$.  These are both Hermitian operators with eigenvalues $\pm 1$.  What is their relationship to each other?  We can get an interesting answer by requiring the walk to satisfy parity symmetry.  Let $\mathcal{P} = \mathcal{P}^\dagger$, $\mathcal{P}^2 = I$ be the {\it parity transformation}. Its action on the physical space is:
\begin{eqnarray}
\mathcal{P} \ket{x} &=& \ket{-x} , \nonumber\\
\mathcal{P} S \mathcal{P} &=& S^\dagger ,
\end{eqnarray}
(where for simplicity we have suppressed the tensor products with the internal space).  Suppose that we require our quantum walk given in Eq.~(\ref{eq:OneDWalk}) to be invariant, $\mathcal{P} U \mathcal{P} = U$, under parity transformations:
\begin{eqnarray}
\mathcal{P} \left[ \left( S  P_R + S^\dagger P_L \right) e^{-i\theta Q} \right] \mathcal{P}
&=& \left( S^\dagger  \mathcal{P} P_R \mathcal{P} + S \mathcal{P} P_L \mathcal{P} \right) \nonumber\\
&& \times e^{-i\theta \mathcal{P} Q \mathcal{P}} \\
&=& \left( S  P_R + S^\dagger P_L \right) e^{-i\theta Q} . \nonumber
\label{eq:Parity1D}
\end{eqnarray}
This implies that $\mathcal{P}$ acts as follows on the internal space operators:
\begin{equation}
\mathcal{P} P_R \mathcal{P} = P_L , \ \ \ 
\mathcal{P} P_L \mathcal{P} = P_R , \ \ \ 
\mathcal{P} Q \mathcal{P} = Q .
\end{equation}
Thus $\mathcal{P}$ anticommutes with $\Delta P$, and commutes with $Q$.  A natural choice for $Q$ is the parity operator $\mathcal{P}$ itself---or, more precisely, the operator that represents the action of $\mathcal{P}$ on the internal space.  But we can make an even stronger statement if we put limits on the dimension of the internal space.

The minimum internal space dimension that allows for two anticommuting operators $\Delta P$ and $\mathcal{P}$ is $d=2$.  In that case, the only nontrivial operators with the properties of $\mathcal{P}$ and $Q$ also satisfy $Q=\pm\mathcal{P}$.  So for $d=2$, the coin operator $Q$ {\it must} be the same as the parity operator.  (We will later see that the same kind of argument applies in three spatial dimensions.)

An immediate result of this is that the walk is {\it unbiased}; that is, that $QP_R Q = P_L$ and $QP_L Q = P_R$.  This is equivalent to anticommutation between $Q$ and $\Delta P$:
\begin{equation}
\left\{ Q, \Delta P \right\} \equiv Q(\Delta P) + (\Delta P)Q = 0 ,
\end{equation}
and implies that $Q$ acts like a ``fair coin flip'' on the internal space.

If we multiply both sides of Eq.~(\ref{eq:Cont1D-3}) by $iQ$ and do some rearrangement, we get the equation
\begin{equation}
i\left( g_0 \partial_t \ket{\psi} + g_1 \partial_x \right)\ket\psi = m \ket{\psi} ,
\label{eq:Cont1D-4}
\end{equation}
where $g_0 \equiv Q$ and $g_1 \equiv Q\Delta P$.  This is a form of the 1D Dirac equation with a two-dimensional internal space.  The operators $g_0$ and $g_1$ anticommute with each other:  $\left\{g_0, g_1\right\} \equiv g_0 g_1 + g_1 g_0 = 0$.  The eigenvalues of $g_0$ are $\pm 1$, and the eigenvalues of $g_1$ are $\pm i$.

The usual Dirac equation in 3D has a four-dimensional internal space; this is the smallest size possible that satisfies the necessary symmetries in three spatial dimensions and one dimension of time.  It is possible, but not necessary, to also use a four-dimensional internal space in one spatial dimension. In that case, however, $Q$ need not be the same as the parity operator.

\subsection{The Momentum Representation}

There is an alternative approach to deriving a relativistic wavefunction in the long-wavelength limit that does not require introducing the spatial difference operator.  This is to work in the momentum representation of the wavefunction, in which the physical space factor in the evolution operator is diagonal (though not the factor that operates on the internal space).

We can define a set of momentum eigenstates as follows.  Consider the eigenvectors of the shift operators $S\ket{x} = \ket{x+\Delta x}$ and $S^\dagger \ket{x} = \ket{x-\Delta x}$.  These take the form
\begin{equation}
\ket{k} = \sum_{j=-\infty}^\infty e^{-i k j\Delta x} \ket{j\Delta x} ,\ \ \ \ -\pi < k\Delta x \le \pi ,
\label{eq:1DMomentumStates}
\end{equation}
which have eigenvalues
\begin{equation}
S\ket{k} = e^{i k\Delta x} \ket{k} ,\ \ \ \ S^\dagger\ket{k} = e^{-i k\Delta x} \ket{k} .
\end{equation}
Obviously, the eigenstates $\{\ket{k}\}$ are not normalizable; but it is possible to rewrite the wavefunction in terms of these eigenstates using the inverse transform
\begin{equation}
\ket{x} = \frac{1}{2\pi} \int_{-\pi}^{\pi} dk\, e^{i k x} \ket{k} ,
\end{equation}
where $x = j\Delta x$ for some integer $-\infty < j < \infty$.

Rewriting the evolution operator (\ref{eq:OneDWalk}) in terms of the momentum basis, it takes a very compact form:
\begin{eqnarray}
U &=& \left( e^{i k\Delta x} P_R + e^{-i k\Delta x} P_L \right) e^{-i\theta Q} \nonumber\\
&=& \left( \cos(k\Delta x) (P_R + P_L) +i\sin(k\Delta x)(P_R - P_L) \right) e^{-i\theta Q} \nonumber\\
&=& \left( \cos(k\Delta x) I +i\sin(k\Delta x) \Delta P \right) e^{-i\theta Q} \nonumber\\
&=& e^{k\Delta x \Delta P} e^{-i \theta Q} .
\label{eq:OneDWalkMomentum}
\end{eqnarray}
In the momentum representation it is straightforward to go to the continuum limit.  It corresponds simply to the long-wavelength limit, $|k| \ll 1$.  Assuming $\theta \ll 1$, as before, we can expand the two exponentials and retain only terms linear in $k$ and $\theta$:
\begin{equation}
\partial_t \ket\psi = i\left( ( k\Delta x/\Delta t) \Delta P - (\theta/\Delta t) Q \right) \ket\psi .
\label{eq:Cont1Dmomentum}
\end{equation}
Transforming back to the position representation and rearranging terms, we recover the two-component Dirac equation in 1D given by Eq.~(\ref{eq:Cont1D-4}).

We will find the momentum representation very helpful when we go to three spatial dimensions, where the symmetries of the equation are much more complicated.

\section{Relativistic Equations in 3D and the Bialynicki-Birula Solution}

The standard form of a quantum walk given in Eq.~(\ref{eq:QWalkEvolution}) can readily be extended to any lattice, or indeed to any regular graph.  This kind of standard quantum walk on, for example, the three-dimensional body-centered cubic (BCC) lattice, would allow the ``walker'' to move from its current vertex to any of 8 neighboring vertices.  This would require an internal space of at least 8 dimensions, to allow for 8 orthogonal projectors $P_j$ corresponding to each of the 8 allowed shifts $S_j$.  It is worthwhile, however, to step back and ask:  what is the {\it smallest} internal space that allows for a quantum walk with nontrivial evolution?

In \cite{Bialynicki94,Bisio15,DAriano17} this question is framed in the following way.  The form of the quantum walk (\ref{eq:QWalkEvolution}) is generalized:
\begin{equation}
\ket{\psi_{t+1}} = U\ket{\psi_t} = \left( \sum_j S_j \otimes A_j \right) \left( I\otimes C \right)\ket{\psi_t} ,
\label{eq:QWalkGeneral}
\end{equation}
where the operators $\{A_j\}$ act on the internal space, but are no longer required to be orthogonal projectors.  The overall unitarity of the walk is no longer guaranteed:  unitarity now requires that the operators $\{A_j\}$ satisfy the condition
\begin{equation}
U U^\dagger = \sum_{j,j'} S_j S^\dagger_{j'} \otimes A_j A^\dagger_{j'} = I ,
\end{equation}
which, since the shifts $\{S_j\}$ are unitary, is equivalent to requiring
\begin{eqnarray}
\sum_j A_j A^\dagger_j &=& I ,\nonumber\\
\sum_{j\ne j'} S_j S^\dagger_{j'} \otimes A_j A^\dagger_{j'} &=& 0 .
\label{eq:UnitarityCondition}
\end{eqnarray}

These conditions in Eq.~(\ref{eq:UnitarityCondition}) were formulated in 1994 in the pioneering work of Bialynicki-Birula \cite{Bialynicki94}, before quantum walks were a field of study.  Bialynicki-Birula was exploring whether it was possible to find discrete versions of the Weyl and Dirac equation.  Using the BCC lattice, he found a solution to (\ref{eq:UnitarityCondition}) with a two-dimensional internal space, which produced the Weyl equation (the Dirac equation with $m=0$).  To find a discrete version of the Dirac equation, he ``pasted together'' two copies of his Weyl evolution with an extra rotation on the internal space.  (A similar method is used in \cite{DAriano17} to derive the Dirac equation in three dimensions as the continuum limit of a quantum walk on the BCC lattice.)  Employing a different method to define the quantum walk, we will show that the condition that the different walk axes and directions be uncorrelated leads to the {\it requirement} that the internal space be the 4-dimensional space of the Dirac gamma matrices. 

The BCC lattice consists of a set of vertices $(x,y,z) = (i\Delta x, j\Delta x, k\Delta x)$, where $i,j,k$ are integers and $\Delta x$ is a fixed lattice spacing.  Two vertices are neighbors if one is $(x,y,z)$ and the other is $(x\pm\Delta x,y\pm\Delta x,z\pm\Delta x)$.  A discrete-time walk on this lattice therefore involves moving one unit $\Delta x$ in each of the three cardinal directions at each timestep. It is this property of the BCC lattice that suggests a different way of defining a quantum walk: by doing three successive steps, one in each direction.  

Employing that approach, we write one step of the 3D walk as:
\begin{eqnarray}
U &=& \left( S_X P^+_X + S_X^\dagger P^-_X \right) e^{-i \theta_X Q_X} \nonumber\\
&& \times \left( S_Y P_Y + S_Y^\dagger P^-_Y \right) e^{-i \theta_Y Q_Y} \nonumber\\
&& \times \left( S_Z P_Z + S_Z^\dagger P^-_Z \right) e^{-i \theta_Z Q_Z} .
\end{eqnarray}
In this equation, $S_{X,Y,Z}$ are shift operators that move the walker one step in the positive $X,Y,Z$ direction; $P^+_{X,Y,Z}$ are projectors that act on the internal space; and $P^-_{X,Y,Z} = I - P^+_{X,Y,Z}$ are the projectors onto their orthogonal complements.  The operators $Q_{X,Y,Z}$ are three ``coin flip'' operators that act on the internal space.  Because we can commute the $e^{iQ}$'s through the projectors (while redefining the projectors themselves), we can without loss of generality move all three coin flip operators to the right side of the expression and absorb them together into a single coin flip:
\begin{eqnarray}
U &=& \left( S_X P^+_X + S_X^\dagger P^-_X \right)
\left( S_Y P^+_Y + S_Y^\dagger P^-_Y \right) \nonumber\\
&& \times \left( S_Z P^+_Z + S_Z^\dagger P^-_Z \right) e^{-i \theta Q} ,
\label{eq:QWalk3DForm}
\end{eqnarray}
where these $P^+$ and $P^-$ operators are not necessarily the same as in the previous expression.  We will use this form of the 3D quantum walk on the BCC lattice for the rest of the paper.  The ordering of $X$, $Y$ and $Z$ in this expression is obviously an arbitrary choice; any other order would work as well to define a 3D quantum walk.

In \cite{Chandrashekar13,Chandrashekar11}, Chandrashekar showed that a quantum walk of the form of Eq.~(\ref{eq:QWalk3DForm}), in addition to being automatically unitary, can give nontrivial dynamics even if the coin flip operator is absent (corresponding to $\theta=0$ in our description here), so long as the projectors on the internal space corresponding to steps in different directions are noncommuting.  As we have noted above (and will see again in the 3D case), the coin flip operator plays the role of the mass term in the continuum limit.  Omitting the coin flip leads to the Weyl equation.

\section{Conditions on Quantum Walks}

We now want to impose conditions on walks of the form (\ref{eq:QWalk3DForm}) analogous to those that arose naturally in the 1D case, and see what happens in the continuum limit.

\subsection{Equal norm condition}

One cannot demand rotational invariance of a walk on a discrete lattice, however one can impose a weaker condition, that the walk not have a preferred lattice axis. Thus we require that it be equally likely for the walker to move in any of the eight allowed directions, or, equivalently, that movement along the different axes should not (necessarily) correlated. In that case a move in the positive X direction, for example, does not bias the move in the positive or negative Y or Z directions. To achieve this, we impose a non-correlation condition on the 3D walk, which we call the {\it equal norm condition}. Surprisingly, as we will see below, this leads to full rotational and boost invariance of the equations of motion in the continuum limit.

To formulate the equal norm condition, we begin by noting that shifts in the positive or negative direction along an axis are determined by the projectors $P^+_{X,Y,Z}$ or $P^-_{X,Y,Z}$, respectively.  Thus if we require $\tr\{P^+_{X,Y,Z}\} = \tr\{P^-_{X,Y,Z}\}$, the walk will not be biased in the positive or negative direction.  We also don't want the shifts along different axes to be correlated.  To avoid this, we require that {\it any} eigenstate of $P^\pm_i$ have equal amplitude in the $P^+_j$ and $P^-_j$ subspaces, for all $i \ne j = X,Y,Z$.  This requirement is equivalent to the following condition on the projectors:
\begin{equation}
P^k_i P^+_j P^k_i = P^k_i P^-_j P^k_i = \frac{1}{2} P^k_i ,
\label{eq:EqualNormCond}
\end{equation}
where
$k = \pm,\ \ i \ne j = X,Y,Z$.

There is a simple geometric interpretation of this condition. Consider a particular space dimension, say, the x-dimension. There are two projection operators associated with that dimension, corresponding to the forward and backward x-directions, and they project onto orthogonal spaces. If the internal space is four-dimensional, for example, each of those spaces is two-dimensional, so they project onto orthogonal planes within the four-dimensional internal space. Similarly, the pairs associated with the y- and z- dimensions project onto orthogonal planes. However the planes associated with x-dimension projection operators are not orthogonal to the planes associated with y- and z-dimension projectors. The equal norm condition requires that all those non-orthogonal planes be oriented at equal angles to each other.  

If we again introduce operators $\Delta P_{X,Y,Z} = P^+_{X,Y,Z} - P^-_{X,Y,Z}$, we can see that the equal norm condition (\ref{eq:EqualNormCond}) is equivalent to requiring that those operators, $\Delta P_{X,Y,Z}$, all anticommute with each other.  To see that, note first that
\begin{equation}
\Delta P_j = P^+_j - P^-_j = 2P^+_j - I = I - 2P^-_j ,
\end{equation}
and that $(\Delta P_j)^2 = I$.  We can then expand out the anticommutator
\begin{eqnarray}
\{\Delta P_i , \Delta P_j \} &=& \Delta P_i \Delta P_j + \Delta P_j \Delta P_i \\
&=& 4\left( P^+_i P^+_j + P^+_j P^+_i \right) - 4\left( P^+_i + P^+_j \right) + 2 I \nonumber\\
&=& 4\left( P^-_i P^-_j + P^-_j P^-_i \right) - 4\left( P^-_i + P^-_j \right) + 2 I . \nonumber
\end{eqnarray}
We can now prove that this anticommutator vanishes for $i\ne j$:
\begin{eqnarray}
\{\Delta P_i , \Delta P_j \} &=& \left(P^+_i + P^-_i\right) \{\Delta P_i , \Delta P_j \} \nonumber\\
&=& P^+_i  \{\Delta P_i , \Delta P_j \} + P^-_i  \{\Delta P_i , \Delta P_j \}\nonumber\\
&=& P^+_i \bigl[ 4\left( P^+_i P^+_j + P^+_j P^+_i \right) \nonumber\\
&& - 4\left( P^+_i + P^+_j \right) + 2 I \bigr] \nonumber\\
&& + P^-_i \bigl[ 4\left( P^-_i P^-_j + P^-_j P^-_i \right) \nonumber\\
&& - 4\left( P^-_i + P^-_j \right) + 2 I \bigr]  \nonumber\\
&=& 4\left( P^+_i P^+_j + P^+_i P^+_j P^+_i \right) \nonumber\\
&& - 4\left( P^+_i + P^+_i P^+_j \right) + 2 P^+_i \nonumber\\
&& + 4\left( P^-_i P^-_j + P^-_i P^-_j P^-_i \right) \nonumber\\
&& - 4\left( P^-_i + P^-_i P^-_j \right) + 2 P^-_i  \nonumber\\
&=& 4\left( P^+_i P^+_j + (1/2) P^+_i \right) \nonumber\\
&& - 4\left( P^+_i + P^+_i P^+_j \right) + 2 P^+_i \nonumber\\
&& + 4\left( P^-_i P^-_j + (1/2) P^-_i \right) \nonumber\\
&& - 4\left( P^-_i + P^-_i P^-_j \right) + 2 P^-_i  \nonumber\\
&=& 0 + 0 = 0 .
\label{eq:ACRProof}
\end{eqnarray}
In the fifth equality we made use of the equal norm condition (\ref{eq:EqualNormCond}), and the rest is just canceling terms.

It is even easier to see that the converse holds.  Suppose that $\{\Delta P_i , \Delta P_j\} = 0$ for all $i\ne j$ and that $(\Delta P_i)^2 = I$ for all $i = X,Y,Z$.  Then
\begin{eqnarray}
P^\pm_i P^+_j P^\pm_i &=& \frac{1}{8} \left(I \pm \Delta P_i \right) \left(I + \Delta P_j \right)
\left(I \pm \Delta P_i \right) \nonumber\\
&=& \frac{1}{8} \biggl[ I \pm 2\Delta P_i \pm (\Delta P_i \Delta P_j + \Delta P_j \Delta P_i) \nonumber\\
&& + (\Delta P_i)^2 + \Delta P_j + \Delta P_i \Delta P_j \Delta P_i \biggr] \nonumber\\
&=& \frac{1}{8} \biggl[ I \pm 2\Delta P_i \pm 0 + I + \Delta P_j - \Delta P_j \biggr] \nonumber\\
&=& \frac{1}{8} \biggl[ 2I \pm 2\Delta P_i\biggr] = \frac{1}{2} P^\pm_i .
\label{eq:ACRConverse}
\end{eqnarray}
This automatically implies
\begin{eqnarray}
P^\pm_i P^-_j P^\pm_i &=& P^\pm_i (I - P^+_j) P^\pm \nonumber\\
&=& P^\pm_i - P^\pm_i P^+_j P^\pm_i \nonumber\\
&=& P^\pm_i - \frac{1}{2} P^\pm_i = \frac{1}{2} P^\pm_i .
\end{eqnarray}

\subsection{Coin flip and the parity operator}

We can generalize the argument based on parity symmetry from the one-dimensional to the three-dimensional case.  Suppose we define a parity operator $\mathcal{P} = \mathcal{P}^\dagger$, $\mathcal{P}^2 = I$ as before, which acts on the position component of the state as follows:
\begin{eqnarray}
\mathcal{P} \ket{x,y,z} &=& \ket{-x,-y,-z} , \nonumber\\
\mathcal{P} S_{X,Y,Z} \mathcal{P} &=& S^\dagger_{X,Y,Z} .
\end{eqnarray}
We now require our three-dimensional quantum walk (\ref{eq:QWalk3DForm}) to be invariant under parity transformations:
\begin{eqnarray}
\mathcal{P}U\mathcal{P} &=& \mathcal{P}\Biggl[ \left( S_X P^+_X + S_X^\dagger P^-_X \right)
\left( S_Y P^+_Y + S_Y^\dagger P^-_Y \right) \nonumber\\
&& \times \left( S_Z P^+_Z + S_Z^\dagger P^-_Z \right) e^{-i \theta Q} \Biggr] \mathcal{P} \nonumber\\
&=& \left( S_X^\dagger \mathcal{P} P^+_X \mathcal{P} + S_X \mathcal{P} P^-_X\mathcal{P}  \right) \nonumber\\
&& \times \left( S_Y^\dagger \mathcal{P} P^+_Y \mathcal{P} + S_Y \mathcal{P} P^-_Y \mathcal{P} \right) \nonumber\\
&& \times \left( S_Z^\dagger \mathcal{P} P^+_Z \mathcal{P} + S_Z \mathcal{P} P^-_Z \mathcal{P} \right)
e^{-i \theta \mathcal{P} Q \mathcal{P}} \nonumber\\
&=& \left( S_X P^+_X + S_X^\dagger P^-_X \right)
\left( S_Y P^+_Y + S_Y^\dagger P^-_Y \right) \nonumber\\
&& \times \left( S_Z P^+_Z + S_Z^\dagger P^-_Z \right) e^{-i \theta Q} ,
\label{eq:Parity3D}
\end{eqnarray}
which implies that
\[
\mathcal{P} P^+_X \mathcal{P} = P^-_X ,\ \ 
\mathcal{P} P^+_Y \mathcal{P} = P^-_Y ,
\]
\[
\mathcal{P} P^+_Z \mathcal{P} = P^-_Z ,\ \ 
\mathcal{P} Q \mathcal{P} = Q .
\]
Once again, we can take the natural choice of $Q$ as the parity operator on the internal space; it then acts as an unbiased coin-flip operator for shifts in all three spatial directions:  $QP^+_{X,Y,Z} Q = P^-_{X,Y,Z}$ and $QP^-_{X,Y,Z} Q = P^+_{X,Y,Z}$.  We immediately know that $Q^2 = I$, and that the eigenspaces of $P^+_{X,Y,Z}$ and $P^-_{X,Y,Z}$ have equal dimension.

We can easily see that the conditions above require that $Q$ anticommutes with each of the $\Delta P_{X,Y,Z}$ operators:
\[
\{ Q, \Delta P_{X,Y,Z} \} = 0 .
\]

We see that imposing both the coin flip condition and the equal norm condition requires us to have four distinct, mutually anticommuting operators:  $Q$, $\Delta P_X$, $\Delta P_Y$, and $\Delta P_Z$.  The minimum dimension which the internal space must have to satisfy this is $d = 4$. As in one space dimension, if we employ the minimum dimension of the internal space, then, aside from a possible minus sign, the parity operator is the only possible choice for Q. 

As we'll see below, in the continuum limit, the coin flip generator $Q$ gives rise to a mass term. In a 1D discrete-time quantum walk, the ``coin space'' and the ``coin flip operator'' were necesarry in order to get nontrivial dynamics.  Without the coin space, the only unitary evolutions possible are steady walking to the right, steady walking to the left, or a completely static walker.  With the coin space, but without the coin flip operator, the evolution in 1D is almost as trivial:  walkers with coins in the $\ket{R}$ state would walk steadily right, while walkers with coins in the $\ket{L}$ state would walk steadily left. When we go to 3D, however, we find that the situation is changed.  

In 3D, a coin space is still necessary for a nontrivial walk, but the coin flip operator is no longer necessary \cite{Chandrashekar13,Chandrashekar11}.  In other words, the particle may have zero mass, but must have nonzero spin.  To be precise, one can obtain nontrivial dynamics by choosing $Q_X = Q_Y = Q_Z = 0$, as long as the projectors $P^+_{X,Y,Z}$ are noncommuting. In that case, the minimal dimension of the coin space is 2, rather than 4. This corresponds to a particle of mass zero, and, as will be clear below, leads to the Weyl equation in the continuum limit.

\subsection{Momentum representation and continuum limit}

Just as in the 1D case (\ref{eq:1DMomentumStates}), we can define momentum states for the 3D walks.  If the position states are $\ket{x,y,z}$ where $x = j\Delta x$, $y = k\Delta x$ and $z = \ell\Delta x$ for some integers $j$, $k$, $\ell$, then the momentum states are
\begin{eqnarray}
&& \ket{k_x,k_y,k_z} \\
&=& \sum_{j,k,\ell=-\infty}^\infty e^{-2\pi i (k_x j\Delta x + k_y k\Delta x + k_z \ell\Delta x} \ket{j\Delta x,k\Delta x,\ell\Delta x} , \nonumber
\label{eq:3DMomentumStates}
\end{eqnarray}
where $-1/2 < k_{x,y,z} \le 1/2$.  These are eigenstates of the shift operators $S_{x,y,z}$ with eigenvalues $e^{2\pi i k_{x,y,z} \Delta x}$.

We can rewrite the time evolution operator in the momentum representation, and we get
\begin{equation}
U = e^{2\pi i k_x\Delta x \Delta P_X} e^{2\pi i k_y\Delta x \Delta P_Y} e^{2\pi i k_z\Delta x \Delta P_Z} e^{-i \theta Q} .
\label{eq:ThreeDWalkMomentum}
\end{equation}
If we go to the limit $|k_{x,y,z}\Delta x| \ll 1$ and $\theta \ll 1$, we get the limiting equation
\begin{eqnarray}
\partial_t \ket\psi &=& \frac{2\pi i\Delta x}{\Delta t} \left( k_x \Delta P_X + k_y \Delta P_Y + k_z \Delta P_Z\right)\ket\psi \nonumber\\
&& - i (\theta/\Delta t) Q \ket\psi .
\label{eq:Cont3Dmomentum}
\end{eqnarray}
Switching back to the position representation, and using the same conventions as in Eq.~(\ref{eq:Cont1D-3}), we get
\begin{equation}
\partial_t \ket{\psi} = c \left( \Delta P_X \partial_x + \Delta P_Y \partial_y + \Delta P_Z \partial_z - im Q \right) \ket{\psi} .
\label{eq:Cont3Dposition}
\end{equation}
This is the Dirac equation in three spatial dimensions.

\subsection{Rotational Invariance}

That the equal norm condition leads to the Dirac equation in the continuum limit means that the walk in that limit is Lorentz invariant (including parity). But it is enlightening to see how, in particular, that discrete symmetry leads to rotation invariance. In the long-wavelength limit $|k_{x,y,z}\Delta x| \ll 1$, we can consider general rotations of the coordinate axes.  Suppose that we apply a rotation to the momentum vector $\vec{k}$:
\begin{equation}
R\vec{k} = \left(\begin{array}{ccc} r_{xx} & r_{xy} & r_{xz} \\
r_{yx} & r_{yy} & r_{yz} \\ r_{zx} & r_{zy} & r_{zz} \end{array} \right)
\left(\begin{array}{c} k_x \\ k_y \\ k_z \end{array} \right)
= \left(\begin{array}{c} k_x' \\ k_y' \\ k_z' \end{array} \right) = \vec{k}' .
\label{eq:MomentumRotation}
\end{equation}
The limiting equation (\ref{eq:Cont3Dmomentum}) will be invariant under such a rotation if the new set of operators $\Delta P_{X,Y,Z}'$ defined by
\begin{equation}
\left(\begin{array}{c} \Delta P_X' \\ \Delta P_Y' \\ \Delta P_Z' \end{array} \right)
= R^T \left(\begin{array}{c} \Delta P_X \\ \Delta P_Y \\ \Delta P_Z \end{array} \right) .
\label{eq:OperatorRotation}
\end{equation}
maintain the required properties.

Do these new $\Delta P_{X,Y,Z}'$ operators have the same properties as the original $\Delta P_{X,Y,Z}$ operators?  They are clearly still Hermitian.  Also, if we look at the square, we get:
\begin{eqnarray}
(\Delta P_X')^2 &=& (r_{xx} \Delta P_X + r_{yx} \Delta P_Y + r_{zx} \Delta P_Z)^2 \nonumber\\
&=& (r_{xx}^2 + r_{yx}^2 + r_{zx}^2) I + r_{xx} r_{yx} \{ \Delta P_X, \Delta P_Y \} \nonumber\\
&& + r_{xx} r_{zx} \{ \Delta P_X, \Delta P_Z \} + r_{yx} r_{zx} \{ \Delta P_Y, \Delta P_Z \} \nonumber\\
&=& I + r_{xx} r_{yx} \{ \Delta P_X, \Delta P_Y \}
+ r_{xx} r_{zx} \{ \Delta P_X, \Delta P_Z \} \nonumber\\
&& + r_{yx} r_{zx} \{ \Delta P_Y, \Delta P_Z \} .
\label{eq:SquareRotation}
\end{eqnarray}
where we have used the property that the columns of the rotation matrix $R$ must be normalized real vectors.  So we retain the property $(\Delta P_X')^2 = I$ because the original operators $\Delta P_{X,Y,Z}$ anticommute.  We get similar equations for $(\Delta P_Y')^2 = I$ and $(\Delta P_Z')^2 = I$. Finally, if we take the anticommutator of two of our new operators, we get
\begin{eqnarray}
\{ \Delta P_X' , \Delta P_Y' \} &=& \{ (r_{xx} \Delta P_X + r_{yx} \Delta P_Y + r_{zx} \Delta P_Z) , \nonumber\\
&& (r_{xy} \Delta P_X + r_{yy} \Delta P_Y + r_{zz} \Delta P_Z) \} \nonumber\\
&=& 2 (r_{xx} r_{xy} + r_{yx} r_{yy}  + r_{zx} r_{zy} ) I \nonumber\\
&& + (r_{xx} r_{yy} + r_{yx} r_{xy}) \{ \Delta P_X, \Delta P_Y \} \nonumber\\
&& + (r_{xx} r_{zy} + r_{zx} r_{xy}) \{ \Delta P_X, \Delta P_Z \} \nonumber\\
&& + (r_{yx} r_{zy} + r_{zx} r_{yy}) \{ \Delta P_Y, \Delta P_Z \} \nonumber\\
&=& (r_{xx} r_{yy} + r_{yx} r_{xy}) \{ \Delta P_X, \Delta P_Y \} \nonumber\\
&& + (r_{xx} r_{zy} + r_{zx} r_{xy}) \{ \Delta P_X, \Delta P_Z \} \nonumber\\
&& + (r_{yx} r_{zy} + r_{zx} r_{yy}) \{ \Delta P_Y, \Delta P_Z \} ,
\label{eq:AnticommutatorRotation}
\end{eqnarray}
where we have used the property that different columns of the rotation matrix $R$ must be orthogonal.  The new operators thus anticommute because the original operators did.  A similar equation applies for $\{ \Delta P_X' , \Delta P_Z' \} = 0$ and $\{ \Delta P_Y' , \Delta P_Z' \} = 0$.  

These results imply a beautiful link between the discrete underlying quantum walk and the continuous long-wavelength limiting equation.  The equal norm condition in the previous subsection, a discrete version of rotational symmetry that was imposed as a noncorrelation requirement on the walk, implies the anticommutation of the three operators $\Delta P_{X,Y,Z}$. The anticommutation relations in turn imply that the equation has full rotational symmetry in the long-wavelength limit. Of course, since the Dirac equation is also invariant under boosts, the limiting equation has full Lorentz symmetry.  

Obviously, for finite $\Delta x$, this Lorentz symmetry is only approximate:  a strong enough boost would violate the condition that $|k_{x,y,z}\Delta x| \ll 1$.  At sufficiently high energies, it would be possible to see the discrete lattice structure underlying the particle's motion; but at lower energies, this discrete structure is completely masked.  It is quite remarkable that by imposing two discrete symmetries on the quantum walk---parity symmetry and the equal-norm condition---we obtain continuous Lorentz symmetry in the long-wavelength limit.

\subsection{The $m=0$ limit}

As mentioned earlier, in three spatial dimensions, a quantum walk of the type we are considering can exhibit nontrivial behavior even without a coin-flip operator.  This is because the projectors $P^\pm_{X,Y,Z}$ can be noncommuting.  In a sense, taking a step in the $X$ direction acts as a coin flip before taking a step in the $Y$ direction, and the step in the $Y$ direction acts as a coin flip before taking a step in the $Z$ direction, which acts as a coin flip before the next step in the $X$ direction, and so forth.

This is in stark contrast to the one-dimensional walk, where a coin-flip operator is necessary to get nontrivial dynamics.  Without it, the particle always moves in the same direction, depending on its initial step.

In taking the long-wavelength limit of the quantum walk in three spatial dimensions, omitting the coin is like taking the limit $m\rightarrow 0$.  The result is the Weyl equation:
\begin{equation}
\partial_t \ket{\psi} = c \left( \Delta P_X \partial_x + \Delta P_Y \partial_y + \Delta P_Z \partial_z \right) \ket{\psi} .
\label{eq:Weyl3D}
\end{equation}
Since the $Q$ operator no longer appears, it is possible to find three Hermitian operators $\Delta P_{X,Y,Z}$ that square to the identity and mutually anticommute with an internal space of only two dimensions.  For example, one could use the $2\times2$ Pauli matrices.

\section{Corrections to the Dirac Equation and Lorentz Violation}

In the continuum limit, the discrete quantum walk that we described is equivalent to the Dirac equation. But if spacetime were discrete, the Dirac equation would be only approximately correct, and the degree to which nature deviates from the Dirac theory, and whether those deviations are observable, would be determined by the size of the lattice spacing $\Delta x$. ($\Delta t$ and $\theta$ are then fixed by the speed of light and particle mass according to $\Delta x/\Delta t = c$ and $\theta = mc\Delta x/\hbar$, respectively.) Experimental tests and astrophysical observations can therefore place upper limits on $\Delta x$.

The most glaring difference between the discrete and continuum theories is that Lorentz invariance is violated in the discrete theory \cite{Arrighi14b,Bisio17}. The possibility that Lorentz invariance violation might play a role in physics was proposed by Dirac himself in the 1950s \cite{Dirac51}, and several others in the years that followed (for example, \cite{Bjorken63,Pavlopoulos67,Redei67}). In the 1990s, influential papers by Coleman and Glashow raised the subject of systematic tests of Lorentz violation within the context of elementary particle physics \cite{Coleman97,Coleman99}. More recently, various theories of quantum gravity have also suggested that Lorentz invariance may not be an exact symmetry \cite{Kostelecky89,Ellis99,Burgess02,Gambini99}. In those theories, the natural scale at which one would expect to observe that violation is at the Planck energy of approximately $10^{19}$ GeV, which would correspond to a lattice spacing of the Planck length, $M_{\rm Pl}=1.616\times10^{-35}$ m. Some suggest that, at smaller scales, the usual notions of space and distance may not even make sense \cite{Carr17}.

The Planck energy is not just far higher than current accelerator energies of approximately $10^3$GeV, but also far higher than the energy of the most energetic observed particles, the Ultra High Energy Cosmic Rays with energies as high as $10^{11}$ GeV $\approx 10^{-8} M_{\rm Pl}$. However, a large violation of Lorentz invariance at the Planck scale can lead to a small degree of violation at much lower energies, presenting the possibility of detection. For example, Lorentz violation can have consequences for neutrino oscillation experiments, in particular that neutrino oscillation still occurs even if the mass is zero \cite{Kostelecky04}. Lorentz violation can also shift the threshold for elementary particle reactions, or lead to the occurrence of other normally forbidden processes such as photon decay and the vacuum Cherenkov effect. As a result, the past few decades have seen a growing literature describing tests of Lorentz invariance, and placing bounds on a variety of proposed deviations (for recent reviews, see \cite{Mattingly05,Liberati13}).

One would expect the nature of proposed Lorentz violations to depend upon the specifics of the Lorentz-violating theory, but one common implication of such theories is an alteration of a particle's dispersion relation, which relates the particle's energy to its momentum and mass \cite{Mattingly05,Liberati13}. That is also one way in which Lorentz violation would show itself in our quantum random walk model.

Though the Dirac theory and quantum electrodynamics, which is based upon it, are two of the most successful theories in all science, the question of whether the Dirac equation or a discrete model is the correct description of nature can only be determined through experiment. The dispersion relation predicted by the quantum walk theory differs from that of the Dirac theory, in that the square of the energy, which is $m^2 c^4 + p^2 c^2$ for the Dirac theory, includes additional terms in the discrete theory, of order $k\Delta x$ and higher, which vanish in the continuum limit.  These higher-order terms can be seen as corrections to the continuum limit, which would act as perturbations to the usual Dirac evolution. These terms have a directional dependence that could make them detectable in a suitably-designed (asymmetric) matter interferometer.  Based on a simple dimensional analysis of the quantum walk studied in this paper, we expect an energy difference between momentum states in different directions to scale like $c p^2 \Delta x/\hbar$.  For thermal neutrons in an interferometer with a length scale of a meter, this could produce a phase shift of order $10^{26} \Delta x$ radians (for $\Delta x$ in meters), which could put very stringent bounds on the lattice spacing $\Delta x$ \cite{BrunMlodinow18b}.


\section{Discussion and Future Work}

In this paper we have shown that a definition of a quantum walk in 3D as three successive 1D walks with the same internal (coin) space leads to a quantum walk on the body-centered cubic lattice, which in the continuum limit gives a Dirac-like equation, where the coin-flip operator gives rise to a mass term.  Imposing discrete symmetry requirements on the quantum walk---in particular, parity symmetry and discrete rotational invariance---requires that the operators in this limiting equation must all mutually anticommute.  This, in turn, means that the internal space must be four dimensional, and that the continuum limit is exactly the Dirac equation, with continuous rotational invariance.

This limiting equation is, of course, Lorentz-invariant; but this symmetry only holds exactly a long length-scales.  There would be small corrections to the Dirac equation, which would show up as direction-dependent perturbations (based on the orientation of the underlying cubic lattice).  This would produce relative phase shifts for wave packets propagating in different directions, which could in principle be detected in matter interferometers, such as interferometry with thermal neutrons.  In a forthcoming paper \cite{BrunMlodinow18b}, we will outline and analyze such an experiment, which could be used either to detect the Lorentz violation, or else to place an upper bound on the granularity of space.

\begin{acknowledgments}

The authors acknowledge helpful conversations with Christopher Cantwell, Yi-Hsiang Chen, Shengshi Pang, Erhard Seiler, and Christopher Sutherland.  They are grateful for the hospitality of Caltech's Institute for Quantum Information and Matter (IQIM).

\end{acknowledgments}


\begin{thebibliography}{99}


\bibitem{AharonovY93} Y. Aharonov, L. Davidovich, and N. Zagury, {\sl Quantum random walks}, Phys. Rev. A {\bf 48}, 1687 (1993).

\bibitem{Ambainis01} A. Ambainis, E. Bach, A. Nayak, A. Vishwanath and J. Watrous, {\sl One-dimensional quantum walks}, in {\sl Proceedings of the ACM Symposium on Theory of Computation (STOCÕ01), 2001} (Association for Computing Machinery, New York, 2001), pp. 37--49.

\bibitem{AharonovD01} D. Aharonov, A. Ambainis, J. Kempe and U. Vazirani, {\sl Quantum Walks On Graphs}, in {\sl Proceedings of the ACM Symposium on Theory of Computation (STOCÕ01), 2001} (Association for Computing Machinery, New York, 2001), pp. 50--59.

\bibitem{Kempe03} J. Kempe, {\sl Quantum walks---an introductory overview}, Contemporary Physics {\bf 44}, 307--327 (2003).


\bibitem{Childs02} A. M. Childs, E. Farhi, and S. Gutmann, {\sl An example of the difference between quantum and classical random walks}, Quant. Inform. Proc. {\bf 1}, 35 (2002).

\bibitem{Shenvi03} N. Shenvi, J. Kempe, and K. Birgitta Whaley, {\sl Quantum random-walk search algorithm}, Phys. Rev. A {\bf 67}, 052307 (2003).

\bibitem{Ambainis03} A. Ambainis, {\sl Quantum walks and their algorithmic applications}, Int. J. Quant. Inf. {\bf 1}, 507 (2003).

\bibitem{Ambainis07} A. Ambainis, {\sl Quantum walk algorithm for element distinctness}, SIAM J. Comput. {\bf 37}, 210 (2007).

\bibitem{Farhi08} E. Farhi, J. Goldstone, and S. Gutmann, {\sl A quantum algorithm for the Hamiltonian NAND tree}, Theo. Comput. {\bf 4}, 169 (2008).

\bibitem{Ambainis10} A. Ambainis, A. M. Childs, B. W. Reichardt, R. \v{S}palek and S. Zhang, {\sl Any AND-OR Formula of Size N Can Be Evaluated in Time $N^{1/2+o(1)}$ on a Quantum Computer}, SIAM J. Comput. {\bf 39}, 2513--2530 (2010).

\bibitem{Reichardt12} B. W. Reichardt and R. \v{S}palek, {\sl Span-program-based quantum algorithm for evaluating formulas}, Theo. Comput. {\bf 8}, 291 (2012).


\bibitem{Moore02} C. Moore and A. Russell, {\sl Quantum walks on the hypercube} in {\sl Proceedings of the 6th International Workshop on Randomization and Approximation Techniques}, 164--178 (Springer-Verlag, London, 2002).

\bibitem{Yin08} Y. Yin, D. E. Katsanos, and S. N. Evangelou, {\sl Quantum walks on a random environment}, Phys. Rev. A {\bf 77}, 022302 (2008).

\bibitem{Brun03a} T.A. Brun, H.A. Carteret and A. Ambainis, {\sl The quantum to classical transition for random walks}, Phys. Rev. Lett. {\bf 91}, 130602 (2003).

\bibitem{Brun03b} T.A. Brun, H.A. Carteret and A. Ambainis, {\sl Quantum random walks with decoherent coins}, Phys. Rev. A {\bf 67}, 032304 (2003).

\bibitem{Kendon03} V. Kendon and B. Tregenna, {\sl Decoherence can be useful in quantum walks}, Phys. Rev. A {\bf 67}, 042315 (2003).

\bibitem{Brun03c} T.A. Brun, H.A. Carteret and A. Ambainis, {\sl Quantum Walks driven by many coins}, Phys. Rev. A {\bf 67}, 052317 (2003).

\bibitem{Kendon06} V. Kendon, {\sl Decoherence in quantum walks---a review}, Math. Struct. in Comp. Sci {\bf 17}, 1169--1220 (2006).


\bibitem{Bialynicki94} I. Bialynicki-Birula, {\sl Weyl, Dirac, and Maxwell equations on a lattice as unitary cellular automata}, Phys. Rev. D {\bf 49}, 6920 (1994).

\bibitem{Meyer96} D. A. Meyer, {\sl From quantum cellular automata to quantum lattice gases}, J. Stat. Phys {\bf 85}, 551 (1996).

\bibitem{Strauch06} F. W. Strauch, {\sl Connecting the discrete- and continuous-time quantum walks}, Phys. Rev. A {\bf 74}, 030301(R) (2006).

\bibitem{Bracken07} A. J. Bracken, D. Ellinas and I. Smyrnakis, {\sl Free-Dirac-particle evolution as a quantum random walk}, Phys. Rev. A {\bf 75}, 022322 (2007).

\bibitem{Chandrashekar10} C. M. Chandrashekar, S. Banerjee and R. Srikanth, {\sl Relationship between quantum walks and relativistic quantum mechanics}, Phys. Rev. A {\bf 81}, 062340 (2010).

\bibitem{DAriano14} G. M. D'Ariano and P. Perinotti, {\sl Derivation of the Dirac equation from principles of information processing}, Phys. Rev. A {\bf 90}, 062106 (2014).

\bibitem{Arrighi14} P. Arrighi, M. Forets and Vincent Nesme, {\sl The Dirac equation as a quantum walk: higher dimensions, observational convergence}, J. Phys. A {\bf 47}, 465302 (2014).

\bibitem{Chandrashekar13} C. M. Chandrashekar, {\sl Two-component Dirac-like Hamiltonian for generating quantum walk on one-, two- and three-dimensional lattices}, Scientific Reports {\bf 3}, 2829 (2013).

\bibitem{DAriano15} G. M. D'Ariano, N. Mosco, P. Perinotti and A. Tosini, {\sl Discrete Feynman propagator for the Weyl quantum walk in 2+1 dimensions}, Europhy. Lett. {\bf 109}, 40012 (2015).

\bibitem{Succi15} S. Succi, F. Fillion-Gourdeau and S. Palpacelli, {\sl Quantum Lattice Boltzmann is a quantum walk}, EPJ Quantum Technology {\bf 2}, 12 (2015).

\bibitem{Farrelly14} T. C. Farrelly and A. J. Short, {\sl Discrete spacetime and relativistic quantum particles}, Phys. Rev. A {\bf 89}, 062109 (2014).

\bibitem{Bisio15} Alessandro Bisio, Giacomo Mauro D'Ariano and Alessandro Tosini, {\sl Quantum Field as a quantum cellular automaton: the Dirac free evolution in one dimension}, Annals of Physics {\bf 354}, 244--264 (2015).

\bibitem{Arrighi15} P. Arrighi, S. Facchini and M. Forets, {\sl Quantum walking in curved spacetime}, arXiv:1505.07023.

\bibitem{Bisio16} Alessandro Bisio, Giacomo Mauro D'Ariano, Marco Erba, Paolo Perinotti, Alessandro Tosini, {\sl Quantum walks with a one-dimensional coin}, Phys. Rev. A {\bf 93}, 062334 (2016).

\bibitem{Arrighi16} P. Arrighi and S. Facchini, {\sl Quantum walking in curved spacetime:  $(3+1)$ dimensions, and beyond}, arXiv:1609.00305.

\bibitem{DAriano17} G. M. D'Ariano, {\sl Physics Without Physics: the Power of Information-theoretical Principles}, Int. J. Theor. Phys. {\bf 56}, 97 (2017).

\bibitem{DAriano17b} G. M. D'Ariano, M. Erba and P. Perinotti, {\sl Isotropic quantum walks on lattices and the Weyl equation}, Phys. Rev. A {\bf 96}, 062101 (2017).

\bibitem{Chandrashekar11} C. M. Chandrashekar, {\sl Two-state quantum walk on two- and three-dimensional lattices}, arXiv:1103.2704.


\bibitem{Arrighi14b} P. Arrighi, S. Facchini and M. Forets, {\sl Discrete Lorentz covariance for Quantum Walks and Quantum Cellular Automata}, New J. Phys. {\bf 16}, 093007 (2014).

\bibitem{Bisio17} Alessandro Bisio, Giacomo Mauro D'Ariano, Paolo Perinotti, {\sl Quantum Walks, Weyl equation and the Lorentz group}, Found. Phys. {\bf 47}, 1065--1076 (2017).

\bibitem{Dirac51} P.A.M.~Dirac, {\sl Is there an Aether?}, Nature {\bf 168}, 906--907 (1951).

\bibitem{Bjorken63} J.D.~Bjorken, {\sl A dynamical origin for the electromagnetic field}, Annals of Physics {\bf 24}, 174--187 (1963).

\bibitem{Pavlopoulos67} T.G.~Pavlopoulos, {\sl Breakdown of Lorentz Invariance}, Physical Review {\bf 159}, 1106--1110 (1967).

\bibitem{Redei67} L.B.~R\'edei, {\sl Validity of Special Relativity at Small Distances and the Velocity Dependence of the Muon Lifetime}, Physical Review {\bf 162}, 1299--1300 (1967).

\bibitem{Coleman97} S.~R.~Coleman and S.~L.~Glashow, {\sl Cosmic ray and neutrino tests of special relativity}, Phys. Lett. B {\bf 405}, 249--252 (1997).

\bibitem{Coleman99} S.~R.~Coleman and S.~L.~Glashow, {\sl High-energy tests of Lorentz invariance}, Phys. Rev. D {\bf 59}, 116008 (1999).

\bibitem{Kostelecky89} V.~A. Kosteleck\'y and and S. Samuel, {\sl Spontaneous Breaking Of Lorentz Symmetry In String Theory}, Phys. Rev. D {\bf 39}, 683 (1989).

\bibitem{Ellis99} J.~Ellis, N.E.~Mavromatos and D.V.~Nanopoulos, {\sl Probing models of quantum space-time foam}, eprint gr-qc/9909085 (1999).

\bibitem{Burgess02} C.P.~Burgess, J.~Cline, E.~Filotas, J.~Matias and G.D.~Moore, {\sl Loop-generated bounds on changes to the graviton dispersion relation}, J. High Energy Phys. {\bf 2002(03)}, 043 (2002).

\bibitem{Gambini99} R.~Gambini and J.~Pullin, {\sl Nonstandard optics from quantum spacetime}, Phys. Rev. D {\bf 59}, 124021 (1999).

\bibitem{Carr17} B.~Carr and S.~B.~Giddings, {\sl Quantum Black Holes}, Scientific American {\bf 292(5)}, 48--55 (2017).

\bibitem{Kostelecky04} V.~A. Kosteleck\'y and A.~Mewes, {\sl Lorentz and CPT violation in the neutrino sector}, Phys. Rev. D {\bf 70}, 031902(R) (2004). 

\bibitem{Mattingly05} D.~Mattingly, {\sl Modern Tests of Lorentz Invariance}, Living Reviews in Relativity {\bf 8}, 5 (2005).

\bibitem{Liberati13} S.~Liberati, {\sl Tests of Lorentz Invariance: a 2013 Update}, Classical and Quantum Gravity {\bf 30}, Number 13 (2013).

\bibitem{BrunMlodinow18b} Todd A. Brun and L. Mlodinow, {\sl Detection of discrete spacetime by matter interferometry}, in preparation.

\end{thebibliography}

\end{document}